\documentclass[a4paper,twoside]{article}
\newcommand{\affil}[1]{$^{\rm #1}$}
\usepackage[authoryear]{natbib}
\bibpunct{(}{)}{;}{a}{}{,}
\usepackage{verbatim}
\usepackage{setspace}
\usepackage{txfonts}
\usepackage{longtable}
\usepackage{balance}
\usepackage[figuresright]{rotating}
\usepackage[a4paper,breaklinks]{hyperref}
\usepackage{natbib} 
\usepackage{graphicx}
\usepackage{latexsym}
\usepackage[utf8]{inputenc}
\usepackage{epsfig}
\usepackage{fancyhdr}
\usepackage{geometry}
\usepackage{tocbibind}
\usepackage{verbatim}
\usepackage{multicol}

\DeclareRobustCommand{\ion}[2]{%
\relax\ifmmode
\ifx\testbx\f@series
{\mathbf{#1\,\mathsc{#2}}}\else
{\mathrm{#1\,\mathsc{#2}}}\fi
\else\textup{#1\,{\mdseries\textsc{#2}}}%
\fi}

\newcommand{\loggf }{\ensuremath{\log\,gf}}
\newcommand{\mlp }{\ensuremath{\alpha_{\mathrm{MLT}}}}
\newcommand{\LHD }{{\sf LHD}}
\newcommand{\xx }{\ensuremath{\mathrm{1D}_{\mathrm{LHD}}}}
\newcommand{\cobold }{{\sf CO$^5$BOLD}}
\newcommand{\mD }{\ensuremath{\left\langle\mathrm{3D}\right\rangle}}
\newcommand{\linfor }{{\sf Linfor3D}}

\def\apj{ApJ}%
\def\apjl{ApJ}%
\def\aap{A\&A}%
\def\solphys{Sol.~Phys.}%
\def\physrep{Phys.~Rep.}%
\date{} 
\title{\large\bf\flushleft The Solar Photospheric Nitrogen Abundance.\\ 
Determination with 3D and 1D model atmospheres.}
\author{\parbox{\textwidth}{\flushleft
\vspace{-0.5cm}
{\it E.Maiorca\affil{A}, E.Caffau\affil{B}, P.Bonifacio\affil{C,B,D}, M.Busso\affil{A,H}, 
R.Faraggiana\affil{E}, M.Steffen\affil{F}, H.-G.Ludwig\affil{C,B}, I.Kamp\affil{G}}\\
\vspace{0.4cm}
{\small \affil{A}\,Dipartimento di Fisica, Universit\`a degli studi di Perugia, via Pascoli, Perugia, I-06123, Italy}\\
{\small \affil{B}\,GEPI, Observatoire de Paris, CNRS, Universit\'e Paris Diderot; 92195
Meudon Cedex, France}\\
{\small \affil{C}\,CIFIST Marie Curie Excellence Team}\\
{\small \affil{D}\,Istituto Nazionale di Astrofisica,
Osservatorio Astronomico di Trieste,  Via Tiepolo 11,
I-34143 Trieste, Italy}\\
{\small \affil{E}\,Dipartimento di Astronomia, Universit\`a 
degli Studi di Trieste, 
via G.B. Tiepolo 11, 34143 Trieste, Italy}\\
{\small \affil{F}\,Astrophysikalisches Institut Potsdam, An der Sternwarte 16, D-14482 Potsdam, Germany}\\
{\small \affil{G}\,Kapteyn Astronomical Institute, Postbus 800, 9700 AV Groningen}\\
{\small \affil{H}\,Istituto Nazionale di Fisica Nucleare, section of Perugia, via Pascoli, Perugia, I-06123, Italy}\\
}}
\begin{document}
\twocolumn[
\begin{changemargin}{.8cm}{.5cm}
\begin{minipage}{.9\textwidth}
\vspace{-1cm}
\maketitle
%
%
\small{\bf Abstract:\\
We present a new determination of the solar nitrogen abundance making use of 3D hydrodynamical 
modelling of the solar photosphere, which is more physically motivated
than traditional static 1D models. We selected suitable atomic 
spectral lines, relying on 
equivalent width measurements already existing in the literature.
For atmospheric modelling we used the \cobold\ 3D radiation hydrodynamics code. 
We investigated the influence of both deviations from local thermodynamic 
equilibrium ("non-LTE effects") and photospheric inhomogeneities ("granulation effects")
on the resulting abundance.
We also compared several atlases of solar flux and centre-disc intensity presently available. 
As a result of our analysis, the photospheric solar nitrogen abundance
is $A(N)$=$7.86\pm0.12$. }


\medskip
\medskip
\end{minipage}
\end{changemargin}
]
\small

\section{Introduction}

\begin{figure*}[ht]
\centering
\mbox{\includegraphics[bb=50 42 575 550, width= 0.48\textwidth]{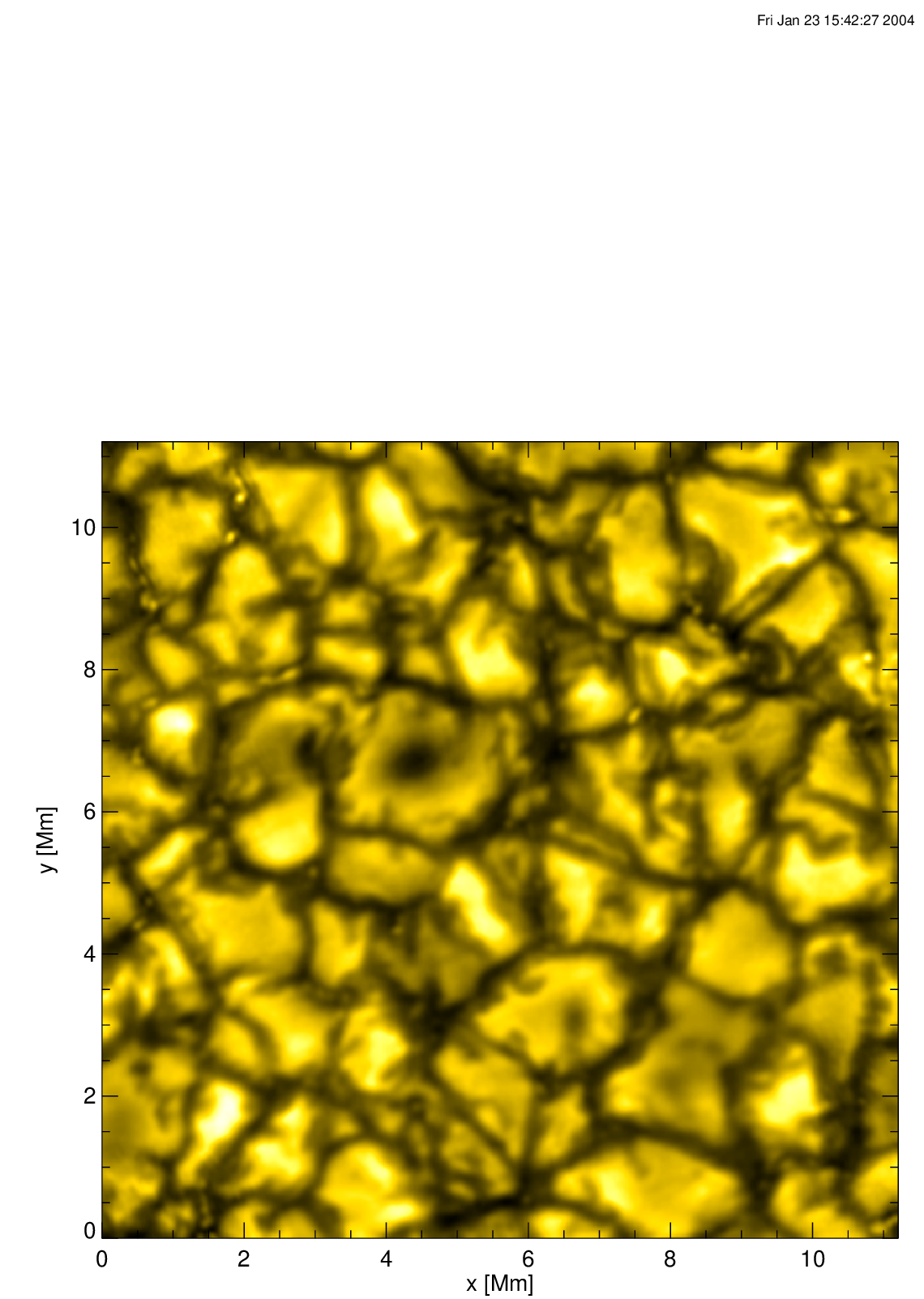}}
\mbox{\includegraphics[bb=50 70 575 578, width= 0.48\textwidth, clip=TRUE]{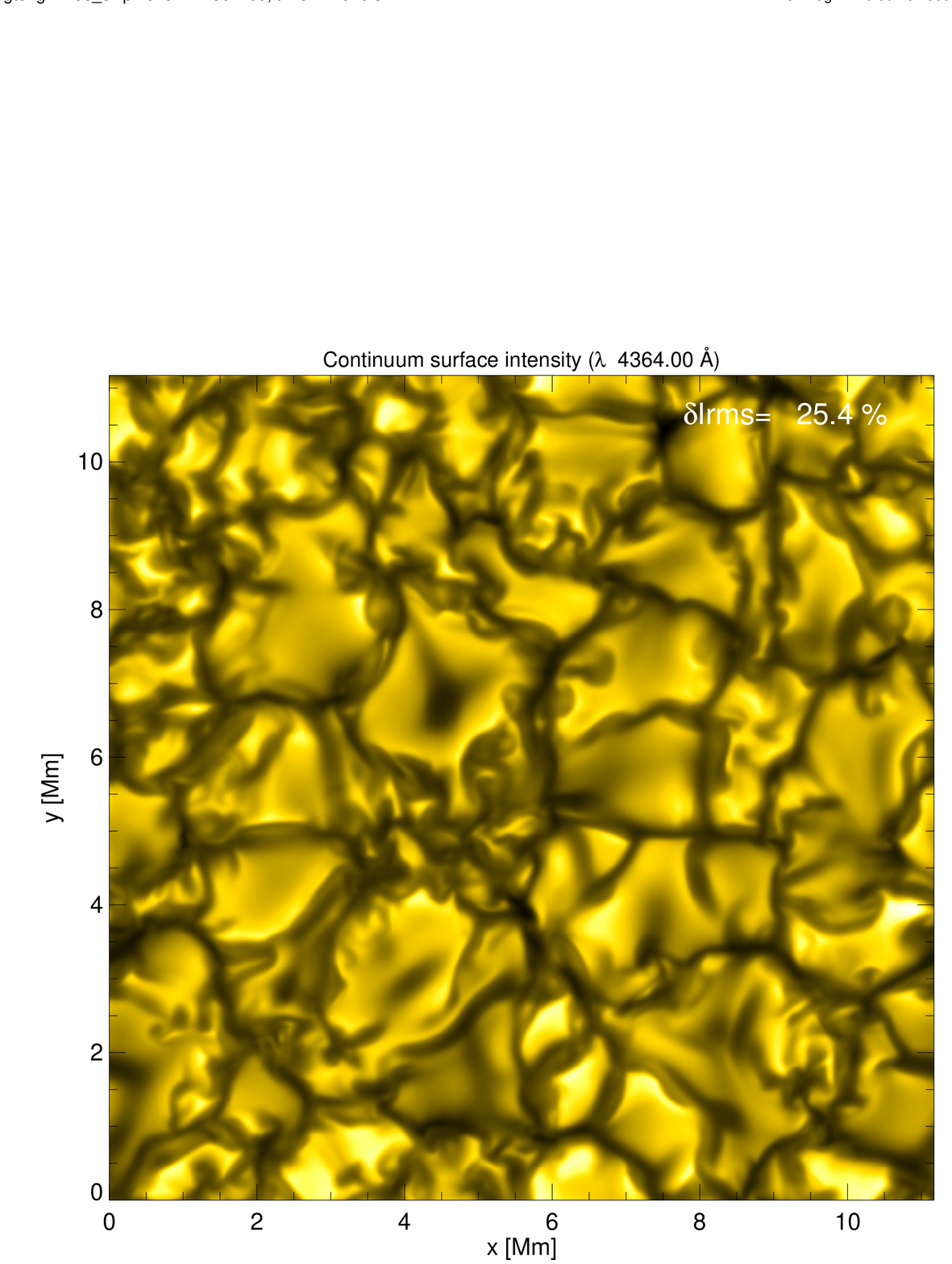}}
\caption{{\bf Left:} Quiet solar granulation as observed with the 1m Swedish Solar
          Telescope (courtesy Mats Carlsson 2004). {\bf Right:} High-resolution
          CO$^5$BOLD simulation of solar surface convection. Both
          images show the emergent continuum intensity (using identical scaling)
          at $\lambda\, 4364$~\AA\ in a field measuring $15^{\prime\prime} \times
          15^{\prime\prime}$ ($11 \times 11$~Mm).}
\label{yd}
\end{figure*}

Images of the Sun reveal a very complex surface structure,
which is referred to as granulation, and may be understood
as the signature of the convective motions in the photosphere.
Traditional static 1D model photospheres ignore all this
complex phenomenology. In the last ten years hydrodynamical
simulations of stellar photospheres have considerably improved
and are now at the stage that they can be compared realistically
against observations. This class of models (here and after referred
to as ``3D models'') is physically better motivated, although computationally
considerably more demanding, than traditional static 1D models.
For the Sun, the comparison of present 3D models with observations 
shows encouraging agreement (e.g.\ Figure~\ref{yd}). The application of 3D
models for abundance work is a largely unexplored territory, but promising
work is in progress not only for the Sun but also for other solar-type stars. 
We began concentrating on chemical abundance determinations based
on the analysis of high resolution spectra and the use of 3D models.
The solar abundances clearly occupy a prominent place in this project,
since we are able to obtain spectra of very high resolution
and S/N ratio for the Sun.
The pioneering works in this field by \citet{carlos} and
\citet{asplund04} have led to a substantial 
downward revision of the solar metallicity, which
implies an awkward tension with the helioseismic
measurements \citep[see][and references therein]{basu}.
It is thus not unsurprising that we start our investigation by 
a reassessment of the abundances of the main contributors
to the solar metallicity, $Z$, i.e. oxygen, nitrogen and carbon.  
The present contribution reports on a redetermination of the solar
nitrogen abundance based on \emph{atomic lines}.

In \citet{sunabboasp}, among other elements,
the solar abundance of nitrogen was also considered,
based on a 3D model.
Unfortunately not much information was given as to which
lines are used, oscillator strengths and other details
of the analysis. Both molecular and atomic lines were considered, 
giving an abundance which is $7.73\pm 0.05$ and $7.85\pm 0.08$, 
respectively (including NLTE-corrections).


\section{Models and \\line formation codes}

Our analysis is mainly based on a 3D model computed with the \cobold\
code \citep{Freytag2002AN....323..213F,
Freytag2003CO5BOLD-Manual,Wedemeyer2004A&A...414.1121W}.
Some basic information on the setup of this numerical simulation 
can be found in \citet{oxy}, who used the same 
solar model to determine the solar oxygen abundance.
We just point out the basic differences between 
the new approach using 3D models and the old approach using 1D models:
while 1D models describe a time-independent, hydrostatic atmosphere, 
a 3D model is the result of solving numerically the time-dependent
hydrodynamic equations together with the equation of radiation transport. 
For any instant of time (``snapshot'' in 3D-jargon)
the 3D models give the physical quantities
on a 3D mesh of points in the photosphere.
This allows a more realistic  
description of the atmosphere (see Figure~\ref{yd}), since both
vertical and horizontal fluctuations of physical quantities can be
taken into account. Moreover, the (turbulent) velocity field in the
stellar atmosphere is automatically obtained without the need to 
specify free parameters (like micro- and macroturbulence).
We would like to point out that, although not requiring the free 
parameters needed in 1D computations to adjust the efficiency of 
the convective energy transport and the strength of the turbulent
velocity field, 3D models are characterized by a set of numerical
parameters, e.g. the numerical scheme used for solving the hydrodynamical 
equations, the spatial resolution of the numerical grid, the amount
of artificial viscosity, the number of rays (angles) considered in 
computing the radiation field, and many others. The hope is that
the results become essentially independent of the choice of these 
parameters once the numerical resolution exceeds some critical 
threshold.

Besides the \cobold\ model, we considered also several
1D models for comparison. 
These include the semi-empirical Holweger-M\"uller model 
(\citealt{hhsunmod, hmsunmod}, hereafter HM),
a 1D model computed with the \LHD\ code (see \citealt{zolfito} 
for further details), an ATLAS solar model computed
by F. Castelli\footnote{\scriptsize http://wwwuser.oats.inaf.it/castelli/sun/ap00t5777g44377k1asp.dat},
and several 1D models with solar parameters computed by ourselves with 
version 9 of the
ATLAS code \citep{1993KurCD..13.....K,2005MSAIS...8...14K}
in its Linux version \citep{2004MSAIS...5...93S,2005MSAIS...8...61S},
using the ``NEW'' Opacity Distribution Functions 
\citep[ODFs;][]{2003IAUS..210P.A20C}, and, finally, a 
1D model obtained by temporal and spatial averaging of the 
3D model over surfaces of equal (Rosseland) optical depth, 
which we call \mD\ model. 

The 3D spectrum synthesis computations are all performed with 
\linfor\footnote{\scriptsize http://www.aip.de/$\sim$mst/Linfor3D/linfor\_3D\_manual.pdf},
which can also compute line formation using different kinds of 1D models 
as input.
For comparison, in the case of 1D models we also used the WIDTH 
code for calculating equivalent widths \citep{1993KurCD..13.....K,2005MSAIS...8...14K,2005MSAIS...8...44C,
2005MSAIS...8...61S} and the SYNTHE code in its Linux version
\citep{1993KurCD..18.....K,2005MSAIS...8...14K, 2004MSAIS...5...93S,2005MSAIS...8...61S} for calculating synthetic spectra.

\section{Line selection, atomic and\\ observational data}
In the literature one can find different choices for the \loggf\ 
values (see Table~\ref{table_1}).
We decided in favour of the NIST data, first of all because,
in this database both computed and measured values are critically examined;
second because all elements ara available, and third to be consistent
with the other papers we produced on photospheric solar abundances.

Our chosen line list and oscillator strengths are given in Table~\ref{atomicdata}.

\begin{table*}[t] 
\begin{small}
\caption{Compilation of the \loggf\ values of permitted \ion{N}{i} lines
used by the different authors.}
\label{table_1}
\begin{center}
\begin{tabular}{clcccccccc}
\noalign{\smallskip}\hline\noalign{\smallskip}
$\lambda$ & NIST   & \multicolumn{2}{c}{Lambert~68} & Lambert~78 &  \multicolumn{2}{c}{Bi{\'e}mont} & Grevesse &   Takeda  & Rentzsch
-Holm\\
  (nm)  &        &  CA    &NBS    &        &  L       &  V       &          &                  &       \\
\noalign{\smallskip}\hline\noalign{\smallskip}
744.2  &$ -0.385$B+&$ -0.33 $ &$ -0.45 $&$ -0.33  $&$  -0.387  $&$ -0.463   $&$  -0.411  $&$ -0.386 $&$    -0.573      $ \\
746.8  &$-0.190 $B+&$-0.16  $ &$-0.27  $&$-0.16   $&$-0.171    $&$-0.248    $&$ -0.208   $&$  -0.171$&$  -0.397        $ \\
818.4  &$-0.286 $B+&$ -0.23 $ &$ -0.42 $&$        $&$          $&$          $&$          $&$        $&$                $ \\ 
820.0  &$ -1.001$B+&$       $ &$       $&$        $&$          $&$          $&$ -0.996   $&$  -1.017$&$  -1.090        $  \\    
821.6  &$+0.132 $B+&$+0.13  $ &$-0.01  $&$ +0.13  $&$ +0.146   $&$+0.089    $&$ -0.106   $&$  +0.147$&$  +0.012        $  \\
822.3  &$-0.271 $B+&$       $ &$       $&$        $&$          $&$          $&$ -0.288   $&$  -0.267$&$  -0.390        $  \\
824.2  &$-0.256 $B+&$       $ &$       $&$        $&$          $&$          $&$  -0.260  $&$  -0.262$&$  -0.360        $  \\
859.4  &$-0.334 $B &$-0.32  $ &$ -0.38 $&$  -0.32 $&$  -0.320  $&$ -0.332   $&$          $&$        $&$                $   \\        
862.9  &$+0.075 $B &$ +0.08 $ &$+0.03  $&$  +0.08 $&$ +0.090   $&$ +0.078   $&$  +0.082  $&$  +0.090$&$  +0.069        $  \\
865.5  &$-0.627 $B &$ -0.62 $& $-0.65  $&$ -0.62  $&$-0.603    $&$-0.616    $&$  -0.608  $&$ -0.603 $&$ -0.630         $ \\
868.3  &$+0.087 $B+&$+0.11  $ &$-0.05  $&$+0.11   $&$+0.115    $&$ +0.102   $&$ +0.109   $&$ +0.116 $&$  -0.051        $ \\
870.3  &$-0.321 $B+&$-0.29  $ &$ -0.41 $&$ -0.29  $&$          $&$          $&$          $&$        $&$                $\\      
871.1  &$-0.234 $B+&$-0.18  $ &$-0.34  $&$        $&$          $&$          $&$          $&$        $&$                $ \\ 
871.8  &$-0.336 $B+&$-0.26  $ &$-0.43  $&$-0.26   $&$-0.338    $&$-0.347    $&$-0.344    $&$  -0.337$&$ -0.419         $  \\
904.5  &$+0.439 $B &$       $ &$       $&$        $&$          $&$          $&$ +0.430   $&$  +0.429$&$                $ \\
939.2  &$+0.320 $B &$ +0.31 $ &$ +0.24 $&$ +0.31  $&$ +0.328   $&$ +0.378   $&$+0.354    $&$  +0.328$&$ +0.316         $   \\
1010.5 &$+0.219 $B+&$       $ &$       $&$        $&$          $&$          $&$ +0.220   $&$  +0.234$&$ +0.200         $ \\
1010.8 &$+0.431 $B+&$ +0.39 $ &$ +0.41 $&$ +0.39  $&$ +0.443   $&$ +0.420   $&$ +0.431   $&$  +0.443$&$  +0.403        $  \\
1011.2 &$+0.607 $B+&$ +0.58 $ &$ +0.60 $&$ +0.58  $&$ +0.622   $&$ +0.600   $&$  +0.611  $&$  +0.623$&$  +0.588        $  \\
1011.4 &$+0.768 $B+&$ +0.74 $ &$ +0.76 $&$ +0.74  $&$ +0.778   $&$ +0.755   $&$ +0.766   $&$  +0.778$&$ +0.751         $ \\
1050.7 &$+0.094 $B &$       $ &$       $&$        $&$          $&$          $&$  +0.249  $&$  +0.249$&$  +0.250        $  \\
1052.0 &$+0.010 $B &$       $ &$       $&$        $&$          $&$          $&$  -0.045  $&$  +0.010$&$   -0.040       $ \\
1053.9 &$+0.503 $B &$ +0.52 $ &$ +0.51 $&$        $&$          $&$          $&$  +0.529  $&$  +0.525$&$  +0.530        $   \\
1075.7 &$-0.608 $C+&$       $ &$       $&$        $&$          $&$          $&$   -0.098 $&$  -0.098$&$  -0.080        $   \\
1238.1 &$+0.247 $C+&$       $ &$       $&$        $&$          $&$          $&$   +0.284 $&$  +0.175$&$  +0.320        $  \\
1246.1 &$+0.480 $B &$       $ &$       $&$        $&$          $&$          $&$    +0.463$&$  +0.437$&$  +0.451        $  \\
1246.9 &$+0.629 $B &$       $ &$       $&$        $&$          $&$          $&$    +0.622$&$  +0.622$&$  +0.610        $ \\
\noalign{\smallskip}\hline\noalign{\smallskip}
\end{tabular}
\\
\end{center}
Notes:
In Lambert~68: CA theoretical Coulomb approximation;
NBS (National Bureau of Standards) is the old NIST;
in Bi\'emont: L length formalism, V velocity formalism.
\end{small}
\end{table*}

\subsection*{Results from the line profile fitting \\using different solar atlases}
\begin{figure}[ht]
\begin{center}
\resizebox{\hsize}{!}{\includegraphics[width=0.5\textwidth, clip=true,angle=0]{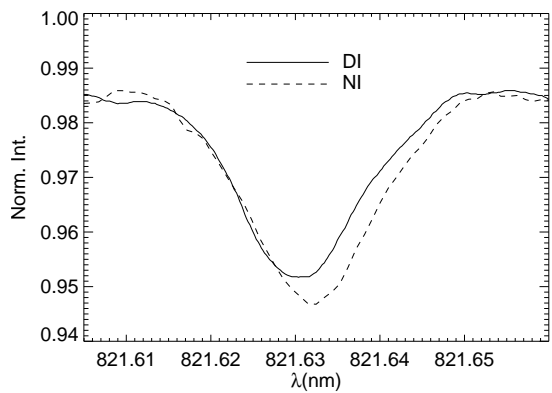}}
\caption{Observed profile of the \ion{N}{i} line at $\lambda$\,821.6\,nm 
as extracted from the Delbouille (solid) and Neckel (dashed) solar 
disc-centre intensity spectral atlases. The differences between the two
solar atlases are not easily explained; possibly telluric absorption
affect the Neckel profile.
}\label{disagr}
\end{center}
\end{figure}

\begin{figure}[ht]
\begin{center}
\resizebox{\hsize}{!}{\includegraphics[width=0.5\textwidth, clip=true,angle=0]{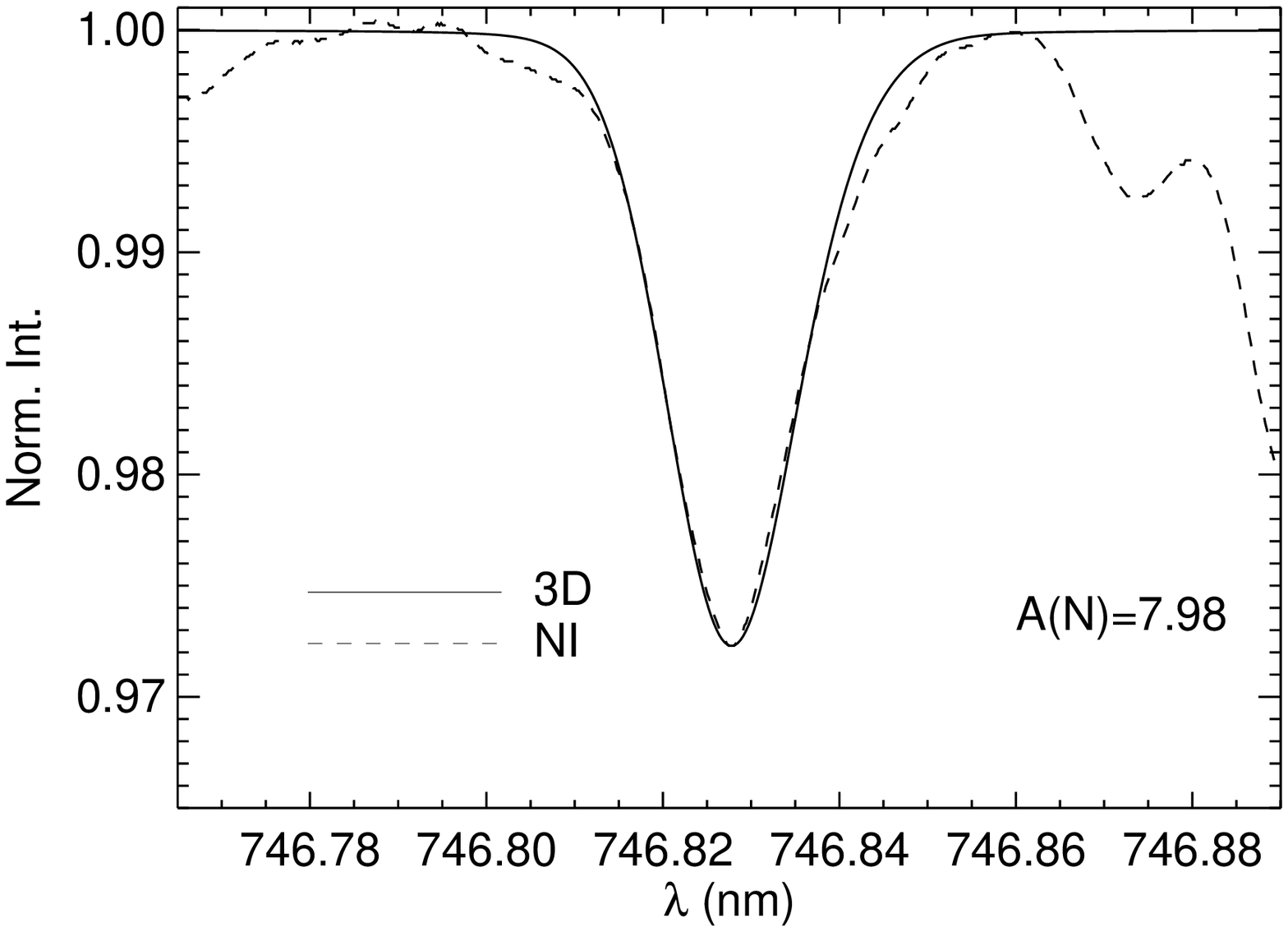}}
\resizebox{\hsize}{!}{\includegraphics[width=0.5\textwidth, clip=true,angle=0]{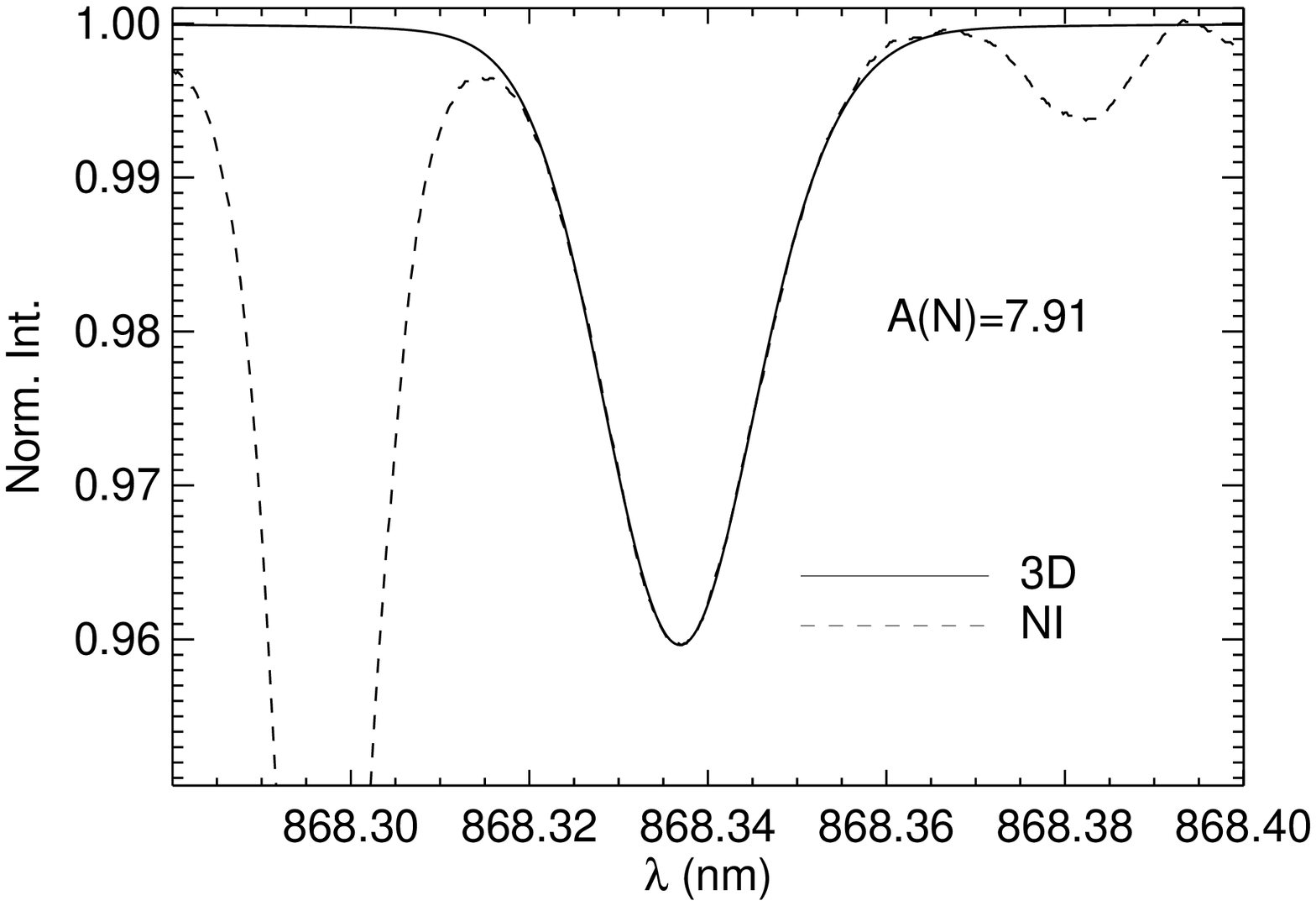}}
\caption{Observed profiles of two \ion{N}{i} lines are compared with a 
3D synthetic profile.
}\label{ni3d}
\end{center}
\end{figure}

We also derived the nitrogen abundance from line profile fitting
for few selected lines.
We did this exercice for comparing the abundance derived from
the four high resolution, high S/N, solar atlases available,
the two centre-disc intensity atlases
(the ``Delbouille'' atlas, i.e. \citealt{delbouille}
and \citealt{delbouilleir} and the 
``Neckel'' intensity atlas, \citealt{neckelobs})
as well as the two solar flux atlases (the ``Kurucz'' 
solar flux atlas, \citealt{kuruczflux} and the ``Neckel''
solar flux atlas, \citealt{neckelobs}).
From previous investigations of the solar observed spectra
(see \citealt{oxy}) we know that these atlases do not always agree.
In Figure \ref{disagr} we can see that the two centre-disk solar atlases 
are not in agreement for the line at $\lambda$\,821.6\,nm.
The solar nitrogen abundance could be deduced from line profile fitting, but
the twelve selected nitrogen lines are blended with molecular and atomic transitions.
We are not sure of the oscillator strength of the blending lines and
we do not have NLTE computations for 
these blending components. 
But for comparing
the results from the four solar atlases this is not a problem.
The four lines we chose are the cleanest, although also for these
lines blends or close-by lines are evident (see Figure\,\ref{ni3d}).
For this reason, being this one only a comparative analysis, 
we performed this exercise only with 1D models, for which the complete line
list with blends is available.

With the ATLAS, HM, and \mD\ solar models as input to SYNTHE,
changing the nitrogen abundance,
we computed three different grids of synthetic spectra.
The fitting code, described in \citealt{caffau05}, is based on a $\chi^2$ minimisation,
and for this purpose uses the MINUIT procedure.


\begin{table}
\caption{Selection of \ion{N}{i} lines in the optical 
and near IR bands}
\scriptsize
\label{atomicdata}
\begin{center}
\begin{tabular}{rccl}
\noalign{\smallskip}\hline\noalign{\smallskip}
$\lambda$ (nm) & E$_{\rm low}$ (eV)  &\loggf  & Q \\ 
\noalign{\smallskip}\hline\noalign{\smallskip}
 744.229  & 10.330 & -0.385 &  B+ \\
 746.831  & 10.336 & -0.190 &  B+ \\
 821.634  & 10.336 & +0.132 &  B+ \\
 822.314  & 10.330 & -0.271 &  B+ \\
 868.340  & 10.330 & +0.087 &  B+ \\
 871.883  & 10.336 & -0.336 &  B+ \\
1010.513  & 11.750 & +0.219 &  B+ \\
1011.248  & 11.758 & +0.607 &  B+ \\
1011.464  & 11.764 & +0.768 &  B+ \\
1050.700  & 11.840 & +0.094 &  B  \\
1052.058  & 11.840 & +0.010 &  B  \\
1053.957  & 11.844 & +0.503 &  B  \\
\noalign{\smallskip}\hline\noalign{\smallskip}
\end{tabular}
\end{center}
\small
Notes:{Error of $\log gf \le 0.03$~dex (Q=B+), $\le 0.08$~dex (Q=B).}
\end{table}
\begin{table}[t]
\caption{A(N) of \ion{N}{i} from line fitting using three different
1D models}
\scriptsize
\label{linefit}
\begin{center}
\begin{tabular}{ccccc|l}
\noalign{\smallskip}\hline\noalign{\smallskip}
Observed & $\lambda$&\multicolumn{3}{c}{A(N)$_{\rm LTE}$}& A(N)$_{\rm LTE}$\\
Spectrum &  (nm)      & HM & ATLAS9 & \mD\ &               Average   \\
\noalign{\smallskip}\hline\noalign{\smallskip}
KF &  746.8 & 7.861 & 7.824 & 7.809 &                      7.873\\
NF &        & 7.874 & 7.837 & 7.827 &                      7.899\\
NI &        & 7.890 & 7.834 & 7.823 &                      7.922\\
DI &        & 7.908 & 7.849 & 7.840 &                      7.883\\
Average &      & 7.883 & 7.836 & 7.825&   \\
Scatter &   & 0.020 & 0.010 & 0.013   &  HM \\
\noalign{\smallskip}\hline\noalign{\smallskip}
KF &  821.6 & 7.847 & 7.811 & 7.805 &                      7.841\\
NF &        & 7.918 & 7.870 & 7.867 &                      7.861\\
NI &        & 7.944 & 7.879 & 7.873 &                      7.864\\
DI &        & 7.808 & 7.752 & 7.746 &                      7.829\\
Average &      & 7.879 & 7.828 & 7.823 &  \\
Scatter &   & 0.063 & 0.059 & 0.060 &  ATLAS9\\
\noalign{\smallskip}\hline\noalign{\smallskip}
KF &  868.3 & 7.910 & 7.889 & 7.857 &                       7.824\\
NF &        & 7.906 & 7.877 & 7.857 &                       7.850\\
NI &        & 7.933 & 7.878 & 7.875 &                       7.857\\
DI &        & 7.932 & 7.886 & 7.882 &                       7.823\\
Average &      & 7.920 & 7.883 & 7.868 &   \\
Scatter &   & 0.014 & 0.006 & 0.013 & \mD\ \\
\noalign{\smallskip}\hline\noalign{\smallskip}
\end{tabular}
\end{center}
\small
Notes:
Col.~(1) is the observed spectrum identification, KF: Kurucz Flux, NF: Neckel Flux,
NI: Neckel Intensity, and DI: Delbouille Intensity;
col.~(2) is the wavelength; col.s~(3)-(5) the A(N)$_{\rm LTE}$ from line profile fitting with
HM, ATLAS9, and \mD\ model, respectively;
col.~(6) average values of the three lines for each atlas, using the model
indicated in the last line of each block.
\end{table}	

Table \ref{linefit} shows that the scatter in the abundance
derived from the 821.6\,nm line from the four solar atlases is considerably 
larger than what is obtained from the other two lines.
The systematic uncertainty due to the choice of a specific solar atlas 
that affects the abundance measurement is computed by comparing the average abundance
obtained considering each atlas. This uncertainty is on average 
$\pm$\,0.02\,dex.
We did not
use 3D synthetic spectra for the fitting procedure. 
In fact, some tens of lines would be necessary
to consider the whole range. 3D run are very time consuming and are,
for the time being, not able to handle too many lines.
Nevertheless, we compared the 3D synthetic profile
to the observed solar spectra. Two examples are visible in Fig.\,\ref{ni3d}.
Only nitrogen is considered in the 3D profile; for this reason the synthetic
profile is not able to reproduce the complete shape of the feature 
(see line at 746.8\,nm) or gives a too high abundance
(see line at 868.3\,nm).

\section{NLTE computations}
The emergent flux depends on the LTE assumption. 
This is a good approximation for lines forming deeply in the photosphere, where the density is high. 
In fact LTE is valid if the photon mean free path is shorter than the distance 
over which matter temperature varies significantly. 
The photon mean free path depends on the probability for the photon to be thermalized, 
hence on the rate of the collisions between the absorbing atoms and electrons or hydrogen. 
This rate increases with matter density.

Since we do not yet have a code able to solve the NLTE problem for nitrogen in the case of a
3D model, we computed the departures from LTE for the \mD\ and the HM model using the Kiel 
code \citep{SH}, with the model atom of \citet{inga96}.
To take into account excitation and ionisation of the nitrogen atoms by
inelastic collisions with neutral hydrogen atoms, the Kiel code uses a 
generalisation of the formalism found in \citet{Drawin}.
This formalism introduces a scaling factor, ${\rm S_H}$, that permits
to modify the efficiency of collisions with hydrogen atoms ($0<{\rm S_H}<1$).
Currently we do not know which value of the scaling factor is the correct one, 
and therefore decided to compute the NLTE-corrections for the two extreme cases 
(${\rm S_H}=0$ and ${\rm S_H}=1)$, and an intermediate value largely used in 
the Kiel community, ${\rm S_H}=1/3$.
It could well be that each nitrogen transition we considered, 
actually require a different value of ${\rm S_H}$. 
Since, in any case, the NLTE-corrections are 
small, we shall consider this differential effect as negligible.

We considered the effects of the horizontal temperature fluctuations on the NLTE-correction. 
The procedure we used is similar to the one that \citet{aufdenberg05} used to
estimate the effects of horizontal temperature inhomogeneities. 
We ordered the emerging flux as a function of temperature. We divided in 12 bins in increasing
temperature and produced horizontally and time-averaged models. 
We computed the NLTE corrections for these twelve average models.

\begin{figure}
\resizebox{\hsize}{!}{\includegraphics[width=0.5\textwidth, clip=true,angle=0]{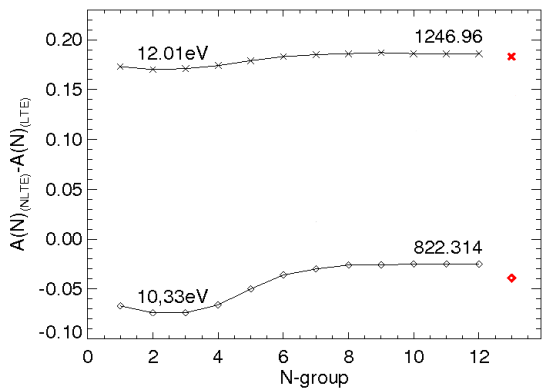}}
\caption{For two representative lines the NLTE corrections
from the twelve group-averaged models, ordered according to increasing
continuum intensity from left to right, are shown together with the
result for the global \mD\ model (rightmost, bold symbol).
The lowest curve is plotted at the true ordinate level, while the other 
is shifted up by 0.20\,dex for clarity.}
\label{grpmod}
\end{figure}

From the results of this 1D-NLTE computation for each line 
we found that the NLTE-corrections are small and not exceeding --0.05\,dex. 
Regarding the effects of the horizontal temperature inhomogeneities, 
we found that lines with lower excitation energy are more sensitive than lines with 
higher excitation energy that are formed deep in the photosphere. 
But these effects are small for our sample of nitrogen lines, since all of them have
high excitation potential (cf.\ Table\,\ref{atomicdata}).
From Figure \ref{grpmod} we can see that the horizontal variation of the NLTE corrections 
is 0.05\,dex at most. For this reason we expect that a full 3D-NLTE 
computation would not differ from our 1D-NLTE calculation by more than 0.03\,dex.
\begin{table*} 
\caption{A(N)$_{\rm LTE}$, A(N)$_{\rm NLTE}$, and 3D-corrections for selected \ion{N}{i} lines 
with EWs from the literature using \linfor}
\scriptsize
\label{ansubsel}
\begin{center}
\begin{tabular}{rcccccccccccc}
\noalign{\smallskip}\hline\noalign{\smallskip}
$\lambda$ &\multicolumn{2}{c}{EW}&\multicolumn{2}{c}{A(N)$_{\rm LTE}$~3D}&
\multicolumn{2}{c}{A(N)$_{\rm NLTE}$~3D}&\multicolumn{2}{c}{A(N)$_{\rm NLTE}$~3D}&
\multicolumn{2}{c}{A(N)$_{\rm NLTE}$~3D}& 3D-\mD & 3D-\xx(\mlp=1.0) \\
    (nm)    & \multicolumn{2}{c}{(pm)} &   &    &\multicolumn{2}{c}{${\rm S_H}=1/3$} & 
              \multicolumn{2}{c}{${\rm S_H}=1$} & \multicolumn{2}{c}{${\rm S_H}=0$}& dex & dex \\
    &     & & & & & & & & &  & & \\
        & G    & B    & G     & B     & G     & B     & G     & B     & G     & B     &        &        \\
 (1)    &  (2) & (3)  & (4)   & (5)   & (6)   & (7)   & (8)   & (9)   & (10)  & (11)  & (12)   & (13)\\
\noalign{\smallskip}\hline\noalign{\smallskip}
  744.2 & 0.26 & 0.27 & 7.808 & 7.826 & 7.774 & 7.792 & 7.782 & 7.800 & 7.770 & 7.788 & $-0.039$ & $-0.002$ \\
  746.8 & 0.52 & 0.49 & 7.961 & 7.931 & 7.923 & 7.893 & 7.931 & 7.901 & 7.919 & 7.889 & $-0.033$ & $+0.008$ \\
  821.6 & 0.86 & 0.87 & 7.854 & 7.860 & 7.802 & 7.808 & 7.817 & 7.823 & 7.790 & 7.796 & $-0.039$ & $+0.007$ \\
  822.3 & 0.24 &      & 7.593 &       & 7.554 &       & 7.565 &       & 7.544 &       & $-0.055$ & $-0.018$ \\
  868.3 & 0.78 & 0.81 & 7.828 & 7.849 & 7.781 & 7.802 & 7.794 & 7.815 & 7.767 & 7.788 & $-0.037$ & $+0.007$ \\
  871.8 & 0.42 & 0.43 & 7.927 & 7.939 & 7.887 & 7.899 & 7.898 & 7.910 & 7.875 & 7.887 & $-0.044$ & $-0.006$ \\
 1010.5 & 0.18 &      & 7.956 &       & 7.939 &       & 7.944 &       & 7.931 &       & $-0.066$ & $-0.020$ \\
 1011.2 & 0.35 & 0.36 & 7.897 & 7.912 & 7.878 & 7.893 & 7.883 & 7.898 & 7.869 & 7.884 & $-0.060$ & $-0.011$ \\
 1011.4 & 0.55 & 0.54 & 7.976 & 7.966 & 7.955 & 7.945 & 7.961 & 7.951 & 7.937 & 7.927 & $-0.053$ & $-0.001$ \\
 1050.7 & 0.14 &      & 8.002 &       & 7.992 &       & 7.995 &       & 7.986 &       & $-0.064$ & $-0.020$ \\
 1052.0 & 0.08 &      & 7.829 &       & 7.819 &       & 7.822 &       & 7.812 &       & $-0.067$ & $-0.024$ \\
 1053.9 & 0.32 &      & 7.989 &       & 7.978 &       & 7.980 &       & 7.971 &       & $-0.057$ & $-0.010$ \\
\noalign{\smallskip}\hline\noalign{\smallskip}
 average&      &      & 7.885 & 7.890 & 7.857 & 7.862 & 7.864 & 7.871 & 7.848 & 7.851 & $-0.051$ & $-0.008$ \\
scatter &      &      & 0.114 & 0.053 & 0.122 & 0.060 & 0.120 & 0.058 & 0.122 & 0.059 &          &          \\
\noalign{\smallskip}\hline\noalign{\smallskip}
\end{tabular}
\end{center}
\small
Notes:
Col.s with G are from \citet{grevesse}, with B are from \citet{biemont}.
Col.~(1) is the wavelength; col.s~(2)-(3) the EWs; col.s~(4)-(5) A(N)$_{\rm LTE}$
from 3D model; col.s~(6)-(11) A(N)$_{\rm NLTE}$ from 3D model
for various values of ${\rm S_{\rm H}}$; col.s~(12)-(13) two different 3D-corrections.
\end{table*}


\section{Nitrogen abundance}
Our final 3D-NLTE nitrogen abundance is obtained by averaging the individual 3D-NLTE 
abundances of each line with equal weight. 
These abundances are obtained with the spectrum synthesis code \linfor\ (see Table \ref{ansubsel}). 
In the same table we show the total 3D-corrections (defined as A(N)$_{\rm 3D}$--A(N)$_{\xx}$) 
and the granulation correction (defined as A(N)$_{\rm 3D}$--A(N)$_{\mD}$) for each line. 
The latter corrections (col.\ (12)) are negative for all lines, indicating that the horizontal 
temperature fluctuations systematically strengthen the (high-excitation) lines. The total 
3D-corrections (col.\ (13)) are systematically more positive, because the \xx\ model produces
slightly stronger lines than the \mD\ model. The 3D-corrections are of the same order of magnitude
as the NLTE-corrections.

Our final result for the solar photospheric nitrogen abundance is:
\begin{equation}
\begin{array}{c @{ A(N)=~} c @{~~~~~{ for} ~~~~~ { S_H} =} c }
     & 7.85\pm 0.12 & 0    \\
     & 7.86\pm 0.12 & 1/3  \\
     & 7.87\pm 0.12 & 1    \\
\end{array}
\label{abundance}
\end{equation}
If only the EWs from \citet{biemont} are considered, 
$A(N)$ is the same while the scatter is reduced to 0.06\,dex.

\subsection*{Discussion}
In theory, the same nitrogen abundance should be derived from each line. 
In practice, this is not the case due to uncertainties in the analysis, 
related to the following elements: the model atmosphere, the \loggf\-values and the EWs.
The results discussed in the following refer to the 3D$_{\rm NLTE}$ model, assuming 
$S_H=1/3$.

Concerning the \loggf\-values and referring to Table \ref{atomicdata}, one could select 
only lines with the Q value equal to B+ and consider the abundances derived only from 
these lines. From the results of Table \ref{ansubsel} we find that selecting only 
lines with $Q=B+$ does not reduce the scatter (both for results from the 3D and the HM 
model) and the mean value becomes $A(N)$=7.83\,dex, very close to our recommended 
value, $A(N)$=7.86\,dex.

The EWs of \citet{biemont} are slightly different from that of \citet{grevesse} by
1 to 6 percent at most. This is an indication that these EWs are reliable. A possible 
way to try to decrease the scatter is to select 
the lines for which we have the EWs both from \citet{grevesse} and \citet{biemont}. 
Since both the two authors chose these lines, we can consider this subset more reliable. 
We find that the mean value does not change, while the scatter for 
the nitrogen abundance from \citet{grevesse} decreases to 0.07\,dex. 
A different possibility is to take the four lines, namely the lines 821.6nm, 871.8nm, 
1011.2nm and 1011.4nm for which EWs of the previous authors  
differ by less than three percent. This agreement 
may be taken to imply that the EWs of these lines are more reliable than the others. 
The results from this subset are $A(N)$=$7.88\pm0.06$ with EWs of \citet{grevesse} 
and $A(N)$=$7.89\pm0.06$ with EWs of \citet{biemont}. From this subset, 
it can be reasonable to discard the line at 1011.4\,nm. In fact
we could not obtain a good fit for this line  and the 
correspondent abundance was too high if compared with the fitting results of the other three lines (see Section 3). 
This behaviour is the same for each model and each atlas we used. 
The nitrogen abundance we obtain from these three lines is
$A(N)=7.86\pm0.05$\,dex with 
EWs of \citet{grevesse}, and $A(N)=7.87\pm0.05$\,dex with EWs of 
\citet{biemont}. With this selection the scatter is further decreased for both sets 
of EWs. 

We conclude from this exercise that the mean nitrogen abundances given 
above are robust against exclusion of the lines which have either
the more uncertain \loggf\-values or exclusion of the lines 
for which we consider the EWs to be less reliable.
 
Our preferred value for the solar nitrogen abundance is
$A(N)$=$7.86\pm 0.12$, assuming the NLTE-correction with $S_H=1/3$.
The indicated error represents the line-to-line scatter and does not
include any uncertainty in the \loggf\-values.


\end{document}